\documentclass[aps,prd,reprint,superscriptaddress,nofootinbib]{revtex4-1}

\usepackage{amsmath}
\usepackage{amssymb}
\usepackage{bbm}
\usepackage{hyperref}
\usepackage{graphicx}
\usepackage{hepnames}
\usepackage{lineno}
\usepackage{subcaption}
\usepackage{mathrsfs}
\usepackage{physics}
\usepackage{upgreek}
\usepackage{xspace}
\usepackage{cleveref}
\usepackage[disable]{todonotes}
\usepackage{booktabs}
\usepackage[separate-uncertainty=true]{siunitx}
\usepackage{cuted}

\graphicspath{{images/}}

\newcommand{\compass}{\textsc{Compass}\xspace} 
\newcommand{\decay}[2]{\ensuremath{#1\!\!\!\!~\to~\!\!#2}}
\newcommand{\Dwave}{\ensuremath{\text{D}}\xspace}
\newcommand{\Fwave}{\ensuremath{\text{F}}\xspace}

\newcommand{\oneMM}{\ensuremath{1^{--}}\xspace}

\newcommand{\Pwave}{\ensuremath{\text{P}}\xspace}

\newcommand{\Swave}{\ensuremath{\text{S}}\xspace}

\newcommand{\free}[1]{\ensuremath{[\Ppi\Ppi]_{#1}}\xspace}
\newcommand{\freed}[2]{\ensuremath{\free{#1}\, \Ppi\, #2}\xspace}

\newcommand{\mdFitAmp}{\ensuremath{\alpha}\xspace}
\newcommand{\miExpAmp}{\ensuremath{\epsilon}\xspace}
\newcommand{\miFitAmp}{\ensuremath{\omega}\xspace}
\newcommand{\miTrueAmp}{\mdFitAmp}
\newcommand{\zmAmp}{\ensuremath{\tilde\eta}\xspace}
\newcommand{\zmVec}{\ensuremath{z}\xspace}

\newcommand{\prodDelta}{\ensuremath{\Delta^{\text{pr}}}}

\DeclareRobustCommand{\Pfzero}{\HepParticle{f}{0}{}\xspace}
\DeclareRobustCommand{\Pftwo}{\HepParticle{f}{2}{}\xspace}
\DeclareRobustCommand{\Pxi}{\HepParticle{\xi}{}{}\xspace}
\DeclareRobustCommand{\PX}{\HepParticle{X}{}{}\xspace}

\newcommand{\Phs}[1]{\ensuremath{\Ph_{#1}}}
\newcommand{\PWAscr}[1]{\ensuremath{\mathcal{#1}}}

\begin{document}

\title{Ambiguities in model-independent partial-wave analysis}

\author{F.~Krinner}
\author{D.~Greenwald}
\affiliation{Technische Universit\"at M\"unchen}
\author{D.~Ryabchikov}
\affiliation{Technische Universit\"at M\"unchen}
\affiliation{IHEP Protvino}
\author{B.~Grube}
\author{S.~Paul}
\affiliation{Technische Universit\"at M\"unchen}

\begin{abstract}

Partial-wave analysis is an important tool for analyzing large data
sets in hadronic decays of light and heavy mesons. It commonly relies
on the isobar model, which assumes multihadron final states originate
from successive two-body decays of well-known undisturbed intermediate
states. Recently, analyses of heavy-meson decays and diffractively
produced states have attempted to overcome the strong model
dependences of the isobar model. These analyses have overlooked that
model-independent, or freed-isobar, partial-wave analysis can
introduce mathematical ambiguities in results. We show how these
ambiguities arise and present general techniques for identifying their
presence and for correcting for them. We demonstrate these techniques
with specific examples in both heavy-meson decay and pion--proton
scattering.

\end{abstract}

\maketitle

\section{Introduction}
\label{sec:introduction}

In hadron spectroscopy, physicists precisely determine the masses,
widths, and other parameters of light mesons and search for new
mesonic states, often in very faint signals
\cite{PhysRevD.95.032004}. In analyzing multibody decays of heavy
mesons (for example, \PB, \PD, and heavy quarkonia), physicists use
spectroscopic techniques to measure both strong and weak phases,
allowing for measurement of CP asymmetries.

In a fixed-target scattering experiment, interaction with a target
excites an incoming particle into a superposition of states that
decays to a set of final-state mesons. All resonant states with
quantum numbers allowed by the conservation laws of the initial
interaction contribute to the superposition. In heavy meson decay,
there is only one decaying state---the heavy meson itself. Both light-
and heavy-meson spectroscopy commonly use partial-wave
analysis~(PWA)---often referred to as Dalitz-plot analysis---which
expands the amplitude for the production of the final state into a sum
of contributions from partial waves: one for each possible combination
of quantum numbers for all states. Each contribution factorizes into
components whose forms are dictated by quantum
mechanics---spin-dependent amplitudes---and components whose forms are
not---dynamic amplitudes, or so-called mass-dependent line shapes. The
dynamic amplitudes parameterize the dependence of the partial-wave
amplitude on the masses of intermediary states.

Dynamic amplitudes are commonly decomposed into sums of contributions
from known resonances. For example, for two pions in a state of spin,
parity, and charge-conjugation parity\footnote{We denote these quantum
  number throughout in this order.} \oneMM with total isospin 1, the
dynamic amplitude can be a sum of contributions from $\Prho(770)$ and
$\Prho'(1450)$. Each resonance has its own dynamic amplitude
model---for example the Breit-Wigner line shape---and an accompanying
complex constant parameterizing its admixture into the total dynamic
amplitude. This decomposition is commonly called the isobar model,
with the individual resonances called isobars.

The results of a partial-wave analysis are strongly dependent on the
quality of the analysis model: for example, on the assumptions of what
resonances to include; how to model their dynamic amplitudes; and what
parameter values to use in those models. PWA with the isobar model
suffers other problems: Models quickly become very complicated when we
include all possible quantum numbers and all known resonances---even
for the production of only three final-state particles. Resonances
with identical quantum numbers that significantly overlap in
mass---like the $\Prho(770)$ and $\Prho'(1450)$---often lead to
unphysical modeling. And dynamic amplitude models often ignore the
strong interactions that can occur between resonances and other
particles in the decay.

These assumptions lead to problems increasingly present in the
analyses of the large data sets provided by current and recent
experiments. In section~\ref{sec:freed_isobar_pwa}, we present a
method to determine dynamic amplitudes directly from the data without
models. Several heavy-meson analyses have used this technique in
recent years \cite{PhysRevD.95.032004, PhysRevD.73.032004,
  PhysLettB.653.1, PhysRep.639.1} and often refer to it as
model-independent PWA---we call it \emph{freed-isobar} analysis. We
demonstrate its applicability to scattering analyses. In
section~\ref{sec:zero_mode}, we demonstrate there exist potentially
fatal mathematical ambiguities that have not been pointed out in
previous analyses.  And in section~\ref{sec:resolving_ambiguities}, we
present several ways to resolve these ambiguities in both heavy-meson
and scattering contexts.

\section{Model-independent partial-wave analysis}
\label{sec:freed_isobar_pwa}

We can remove model dependencies of the isobar model by determing
dynamic amplitudes from the data. Instead of decomposing a dynamic
amplitude into contributions from intermediary resonances---each with
its own dynamic amplitude model---we parameterize it as a complex step
function.

Before detailing the formalism of these freed isobars, we briefly
review the standard PWA formalism using the isobar model. To make some
of the formulas concrete and to simplify notation, we refer to the
example of three-pion production, from both pion--proton scattering,
\begin{equation}
  \label{eqn:pi_p_scattering}
  \Ppiminus \Pproton \to \Ppiminus\Ppiplus\Ppiminus \Pproton,
\end{equation}
and \PD meson decay,
\begin{equation}
  \label{eqn:D_3pi_decay}
  \decay{\PDminus}{\Ppiminus\Ppiplus\Ppiminus}.
\end{equation}
A fully generic description of the isobar model is available in
reference~\cite{PhysRevD.11.3165}.

\subsection{Partial-wave analysis}
\label{subsec:pwa}

In partial-wave analysis, we assume events are distributed according
to the square of the sum of the partial-wave amplitudes---the
intensity of the model. Each amplitude describes a unique transition
through intermediary states with well-defined quantum numbers to the
final state. We assume the transition proceeds through two-body
decays. In our three-pion examples, the quantum numbers defining a
partial wave are the spin ($J$), spin projection~($M$), parity~($P$),
and charge-conjugation parity~($C$) of the three-pion system,~\PX; the
spin of the intermediary two-pion system~($S$); and the angular
momentum~($L$) between the two-pion system and the third pion,
referred to as the spectator pion. Since there are two possible
combinations of pions to form the intermediary system, the sum over
amplitudes also sums over the two combinations---this is commonly
called Bose symmetrization, with each combination referred to as a
symmetrization.

Each partial-wave amplitude has a spin-dependent amplitude component,
$\psi(m_{3\Ppi}, \vec\tau)$, which is dependent on the mass of the
initial system, $m_{3\Ppi}$, and the coordinate in phase space,
$\vec\tau$, of the final-state particles.\footnote{For three-particle
  decay, one needs five independent coordinates to specify
  $\vec\tau$. For scattering, these are usually two Gottfried-Jackson
  angles~\cite{NuovoCim.33.309}; two angles, each one between two
  final-state momenta; and the invariant mass of one pair of
  final-state particles. For heavy-meson decay, these are usually two
  invariant masses of pairs of final-state particles; and three Euler
  angles describing the overall orientation of the decay plane. If the
  heavy meson is spinless, as in our example, the Euler angles can be
  omitted since the decay is isotropic in them.} This component is
fully specified by the quantum numbers of the wave; but its exact form
is formalism dependent.

Each partial wave also has a dynamic amplitude for the production of
the wave, $\prodDelta(m_{3\Ppi})$, and one for the production of the
intermediary state, $\Delta(m_{2\Ppi})$. Each dependends only on the
mass of the state whose production it parameterizes---not on any other
phase-space coordinates. Commonly, one also includes angular-momentum
barrier factors and form factors, which are dependent on both
$m_{3\Ppi}$ and $m_{2\Ppi}$---we denote the product of such terms by
$F(m_{3\Ppi},m_{2\Ppi})$.

The total amplitude, summing over partial waves, $a$, and
symmetrizations is
\begin{widetext}
  \begin{equation}
    \label{eqn:pwa_amplitude}
    \PWAscr{A}(m_{3\Ppi}, \vec\tau) \equiv
    \hat{\sum_{a}}\ \prodDelta_a(m_{3\Ppi}) \, \hat\psi_a(m_{3\Ppi},\vec\tau) \, F_a(m_{3\Ppi},\hat{m}_{2\Ppi}) \, \Delta_a(\hat{m}_{2\Ppi}),
  \end{equation}
\end{widetext}
where we denote symmetrization-dependent functions and variables (and
the sum itself) by hats---we use this notation throughout the paper.

For scattering experiments, data are conventionally divided into bins
of~$m_{3\Ppi}$ that are analyzed independently, so that the dynamic
amplitudes for the production of the three-pion states are learned
empirically. In this case, $\prodDelta_a(m_{3\Ppi})$ is a set of
complex parameters, one for each three-pion mass bin,~$b$---which we
call production amplitudes. We rewrite
equation~(\ref{eqn:pwa_amplitude})---now applying independently for
each three-pion mass bin---as a sum over partial waves, each with a
production amplitude, $\Delta^{(b)}_a$:
\begin{equation}
  \label{eqn:scattering_amp}
  \PWAscr{A}^{(b)}(\vec\tau) \equiv \hat{\sum_{a}}\ \Delta^{(b)}_a \, \hat\psi_a(\vec\tau) \, \Delta_a(\hat{m}_{2\Ppi});
\end{equation}
since within one three-pion mass bin~$F(m_{3\Ppi},\hat{m}_{2\Ppi})$ is
dependent only on the two-pion mass, we absorb it
into~$\Delta_a(\hat{m}_{2\Ppi})$. To simplify notation, since it is
sufficient to discuss the formalism within a single three-pion mass
bin, we omit the mass-bin index in the remainder of the paper.

In heavy-meson decay, there is only one initial state---the heavy
meson itself---with a fixed mass, so we can
parameterize~$\prodDelta_a(m_{3\Ppi})$ by a single complex variable
and also use equation~(\ref{eqn:scattering_amp}).

\subsection{Model-dependent isobars}
\label{sec::model-dep-isobars}

There are many ways to formulate dynamic amplitudes, none of which are
dictated by first principles. The most common way is the isobar model,
in which the dynamic amplitude for the two-pion state is a sum of
contributions from known resonances, \Pxi, with the quantum numbers of
the two-pion state in the wave:
\begin{equation}
  \label{eqn:isobar_model_dynamic_amp}
  \Delta_a(m) = \sum_{\Pxi} \mdFitAmp^a_{\Pxi} \Delta_{\Pxi}(m).
\end{equation}
Each resonance is parameterized by an individual dynamic amplitude,
$\Delta_{\Pxi}(m)$, and a complex admixture variable,
$\mdFitAmp^a_{\Pxi}$. There are myriad ways to formulate the resonance
dynamic amplitudes. One of the most common is the relativistic
Breit-Wigner shape:
\begin{equation}
  \label{eqn:BW}
  \Delta_{\Pxi}^{\text{BW}}(m) \equiv \frac{m_{\Pxi} \Gamma_{\Pxi}}{m_{\Pxi}^2 - m^2 - i m_{\Pxi} \Gamma_{\Pxi}},
\end{equation}
which ascribes a mass, $m_{\Pxi}$, and width, $\Gamma_{\Pxi}$, to the
resonance particle.\footnote{The numerator $m_{\Pxi} \Gamma_{\Pxi}$
  normalizes the shape so that $\Delta_{\Pxi} = i$ at $m = m_{\Pxi}$.}

We can absorb the production amplitude into the $\mdFitAmp_{\Pxi}$,
and rewrite equation~(\ref{eqn:scattering_amp}) as a sum over waves
and the resonances contributing to each wave:
\begin{equation}
  \label{eqn:model_dependent_amp_three-pion}
  \PWAscr{A}(\vec\tau) = \hat{\sum_{a}} \sum_{\Pxi}\, \hat\psi_a(\vec\tau) \, \mdFitAmp_{\Pxi}^{a} \Delta_{\Pxi}(\hat{m}_{2\Ppi}),
\end{equation}
where each complex-valued $\mdFitAmp_{\Pxi}^a$ parameterizes the
production of a resonance and the spectator pion in the total spin
configuration of the wave: for example, a three-pion state with
quantum numbers $0^{-+}$ decaying into $\Pfzero\Ppiminus$ in a
relative \Swave wave, where \Pfzero is a $\Ppiminus\Ppiplus$ resonance
with quantum numbers $0^{++}$ and an assumed dynamic amplitude model
(and parameter values therein); or a three-pion state with quantum
numbers $2^{-+}$ decaying into $\Prho\Ppiminus$ in a relative \Pwave
wave, where the \Prho is a $\Ppiminus\Ppiplus$ resonance with quantum
numbers $1^{--}$ and an assumed dynamic amplitude
model.\footnote{States with spin greater than zero should also have a
  spin projection specified, which we leave off here for brevity.}

The isobar model has been very useful in analysis of scattering
experiments and heavy-meson decays. But it requires assumptions
concerning what resonances are present and what forms their dynamic
amplitudes have. These assumptions may bias analyses and ignore small
structures not easily modeled, which can distort fit results. These
possible substructures---increasingly more visible in the larger and
larger data sets of modern experiments---may arise from new resonant
states or from final-state interactions.

\subsection{Model-independent isobars}
\label{sec::freedPWA}

In freed-isobar PWA, we remove the model dependency inherent in the
isobar model by parameterizing dynamic amplitudes as complex step
functions:
\begin{equation}
  \label{eqn:step_function}
  \Delta_a(m) = \sum_{\beta} \miFitAmp^{a}_\beta \mathbbm{1}_\beta(m),
\end{equation}
where the $\beta$ are disjoint bins of the two-pion mass range;
$\miFitAmp_\beta$, the complex values of the dynamic amplitude in those
ranges; and $\mathbbm{1}_\beta$, the indicator function,
\begin{equation}
  \mathbbm{1}_\beta(m) \equiv
  \begin{cases}
    1, & \text{ if } m \in \beta,\\
    0, & \text{ if } m \notin \beta.
  \end{cases}
\end{equation}
The division of the mass range into bins is independent for each wave,
but is identical for all symmetrizations of a wave. With this
model-independent dynamic amplitude, the description of an isobar is
freed from assumptions on both what resonances comprise it and how to
formulate the dynamic amplitudes of those resonances.

Substituting into equation~(\ref{eqn:scattering_amp}) and absorbing
the production amplitude into the $\miFitAmp^a_\beta$, we have
\begin{equation}
  \label{eqn:model_independent_amp}
  \PWAscr{A}(\vec\tau) = \hat{\sum_a}\sum_{\beta} \hat\psi_a(\vec\tau)\, \miFitAmp^{a}_{\beta}\, \mathbbm{1}_\beta(\hat{m}_{2\Ppi})
\end{equation}
This has a form identical to
equation~(\ref{eqn:model_dependent_amp_three-pion}), with each
two-pion mass bin in each wave appearing as an intermediate state with
an indicator function for a dynamic amplitude. So we can use the same
computational techniques (and software) used for model-dependent PWA.

\section{Zero modes in freed-isobar PWA}
\label{sec:zero_mode}

If we allow dynamic amplitudes for isobars to have abritrary forms,
mathematical ambiguities may arise: there may exist functions,
$\tilde\Delta_a$, whose combined amplitude vanishes---
\begin{equation}
  \label{eqn:zero_mode_functions}
  \hat{\sum_a}\ \hat\psi_a(\vec\tau) \, \tilde\Delta_a(\hat{m}_{2\Ppi}) = 0
\end{equation}
---no matter what values are taken for parameters of the functions,
including an overall scaling of them. Since the amplitude is zero, the
parameters of these functions are superfluous degrees of freedom in
the total PWA amplitude. We refer to each set of $\tilde\Delta_a$ as a
zero mode. For such a zero mode to exist, there must be at least two
terms in the sum of equation~(\ref{eqn:zero_mode_functions}); that is,
there must be at least two symmetrizations or two waves in the sum.

The binned functions of model-independent PWA introduce enough freedom
to the dynamic amplitudes that the sum in
equation~(\ref{eqn:zero_mode_functions}) can be approximately
zero---we can have modes that contribute very weakly to the overall
amplitude. These weakly contributing modes can complicate analyses and
obscure underlying results. Since it is these approximately zero modes
that show up in model-independent PWA, we refer to them also simply as
zero modes when there is no possibility of confusion.

We can decompose the $\miFitAmp^{a}_{\beta}$ of
equation~(\ref{eqn:model_independent_amp}) into a contribution
describing nature, $\miTrueAmp^{a}_{\beta}$, and a contribution from
zero modes:
\begin{equation}
  \label{eqn:mi_amp_components}
  \miFitAmp^{a}_{\beta} = \miTrueAmp^{a}_{\beta} + \sum_{\zmVec} \zmAmp_{\zmVec} \zmVec^{a}_{\beta},
\end{equation}
where each zero mode has complex scaling factor $\zmAmp_{\zmVec}$ and
values, $\zmVec^{a}_{\beta}$, in the two-pion mass bins that
approximate a $\tilde\Delta_{a}(m)$:
\begin{equation}
  \label{eqn:zero_mode_def}
  \tilde\Delta^{\zmVec}_{a}(m) \approx \zmAmp_{\zmVec} \sum_\beta \zmVec^{a}_\beta \mathbbm{1}_\beta(m).
\end{equation}
Without loss of generality we can define a single zero mode as a
collection of real functions with one common complex scaling
factor---so that the $\zmVec^{a}_\beta$ are real.

\subsection{A concrete example}
\label{sec:zero_mode_example_case}

Let us demonstrate the presence of a zero mode in our three-pion
examples: The final state particles are all spinless. Let us also
consider a spinless three-pion state---which is possible in
pion--proton scattering and is always the case in \PD decay. With the
initial-state and final-state particles all spinless, there is only
one spin quantum number to consider. Since there are no doubly-charged
mesons, we consider only the $\Ppiplus\Ppiminus$ intermediary states,
of which there are two symmetrizations. We label waves by the spin of
the $\Ppiplus\Ppiminus$ system; and label the pions as
$\Ppi^-_1\Ppi^+_2\Ppi^-_3$ and the two $\Ppiplus\Ppiminus$
symmetrizations as 12 and 23.

We will fulfill equation~(\ref{eqn:zero_mode_functions}) with functions
in the \Swave and \Pwave waves:
\begin{alignat}{2}
  \label{eqn:generic_s_p_wave_zero_mode}
     & \psi^{\Swave}_{12}(\vec\tau)\,\tilde\Delta^{\Swave}(m_{12})\ +\ \psi^{\Pwave}_{12}(\vec\tau)\,\tilde\Delta^{\Pwave}(m_{12}) &\nonumber\\
  +\ & \psi^{\Swave}_{23}(\vec\tau)\,\tilde\Delta^{\Swave}(m_{23})\ +\ \psi^{\Pwave}_{23}(\vec\tau)\,\tilde\Delta^{\Pwave}(m_{23}) &\ = 0.
\end{alignat}
The \Swave-wave spin-dependent amplitudes are unity. To find explicit
forms for the zero-mode dynamic amplitudes, we must assume a formalism
for the \Pwave-wave spin-dependent amplitudes. We use the Zemach
tensor formalism of~\cite{PhysRev.140.B97} because it is simple and
common:
\begin{equation}
  \label{eqn:Zemach_general}
  \psi^L_{12}(\vec\tau) = |\vec{p}_1|^{L}\ |\vec{p}_3|^L\, P_L\!\left(\hat{p}_1\cdot\hat{p}_3\right),
\end{equation}
where $P_L$ is the $L$'th-order Legendre polynomial and the momenta
are in the $\Ppi_1\Ppi_2$ rest frame. For the \Pwave wave,
\begin{equation}
  \psi^{\Pwave}_{12}(\vec\tau) = \frac{1}{4}\left(m^2_{12} + 2m^2_{23} - m_{3\Ppi}^2 - 3m^2_{\Ppi}\right);
\end{equation}
$\psi^{\Pwave}_{23}$ is formed by swapping $m_{12}$ and $m_{23}$. If
we set $\tilde\Delta^{\Pwave}(m)$ constant, then the \Pwave-wave
contribution in equation~(\ref{eqn:generic_s_p_wave_zero_mode}) has
terms that are either independent of the two-pion masses or dependent
on only one two-pion mass---that is, there are no terms dependent on
both $m_{12}$ and $m_{23}$. We can cancel all these terms with the
\Swave-wave dynamic amplitude. The explicit zero-mode dynamic
amplitudes are
\begin{alignat}{1}
  \tilde\Delta^{\Pwave}(m) &= 4\zmAmp
  \label{eqn:sp_zero_mode_p}\\
  \tilde\Delta^{\Swave}(m) &= \zmAmp(m_{3\Ppi}^2 + 3m^2_{\Ppi} - 3m^2).
  \label{eqn:sp_zero_mode_s}
\end{alignat}
The zero mode has two degrees of freedom, those of the arbitrary
complex coefficient~$\zmAmp$.

In the appendix, we give a fuller picture of zero modes in the decay
of a spinless particle to three spinless particles and show an example
of a zero mode contained entirely in one freed amplitude in a spinful
decay.

\hspace{1\baselineskip}
\subsection{Numerically determining zero modes}
\label{sec:numerical_zero_modes}

Most zero modes are not as simple as the example above. Those for
higher-spin decays are particularly more complicated. But we can
numerically determine their shapes in the freed-isobar formulation.

In model-independent PWA, a zero mode satisfies
\begin{equation}
  \label{eqn:freed_isobar_zero_mode}
  \hat{\sum_{a,\beta}}\ \hat\psi_a(\vec\tau) \, \zmVec^{a}_{\beta}\mathbbm{1}_{\beta}(\hat{m}_{2\Ppi}) \approx 0,
\end{equation}
where the $\zmVec^{a}_{\beta}$ are the values in each two-pion mass
bin of each freed wave such that the sum is very small---they are real
since we have defined a zero mode as real in
equation~(\ref{eqn:zero_mode_def}). The standard mathematical tool to
solve for the $\zmVec^{a}_{\beta}$ that fulfill
equation~(\ref{eqn:freed_isobar_zero_mode}) is to look for the
eigenvectors of the Gram matrix of the freed isobars that have
vanishingly small eigenvalues. This matrix is
\begin{widetext}
  \begin{equation}
    \label{eqn:normalized_integral_matrix}
    \pmb{I}_{a\beta,b\delta} \equiv
    \frac{1}{\PWAscr{N}_{a\beta}\,\PWAscr{N}_{b\delta}}
    \int\!\! 
    \qty(\hat\sum\, \hat\psi^*_a(\vec\tau) \mathbbm{1}_\beta(\hat{m}_{2\Ppi}))
    \qty(\hat\sum\, \hat\psi_b(\vec\tau) \mathbbm{1}_{\delta}(\hat{m}_{2\Ppi}))
    \dd\vec\tau
  \end{equation}
\end{widetext}
where $a$ and $b$ label waves, $\beta$ and $\delta$ label
two-pion-mass bins in each, respectively, and the sums are over
possible symmetrizations of each wave. The normalization constants
\begin{equation}
  \PWAscr{N}^2_{a\beta} \equiv \int
  \abs{\hat\sum\, \hat\psi_a(\vec\tau) \mathbbm{1}_\beta(\hat{m})}^2 
  \dd\vec\tau,
\end{equation}
are chosen such that the diagonal elements of the matrix are unity.
They mitigate spurious effects from two-pion mass bins that hang over
an edge of phase space. It is enlightening to connect this
mathematical tool back to a physical interpretation: This is the
overlap integral matrix for our two-pion mass bins. An eigenvector of
it with a very small eigenvalue is a set of dynamic amplitude values
in each bin of each wave that contribute neglibly to the overall
intensity. The set of freed-isobar dynamic amplitudes forming a zero
mode are therefore
\begin{equation}\label{eqn:binned_zero_mode_def}
  \tilde\Delta^{z}_a(m) = \zmAmp_\zmVec \sum_{\beta} \zmVec^a_\beta\, \mathbbm{1}_\beta(m),
  \qquad
  \zmVec^{a}_{\beta} \equiv \tilde\zmVec^{a}_\beta \PWAscr{N}_{a\beta}^{-1}
\end{equation}
where $\tilde\zmVec^{a}_{\beta}$ are the elements of the eigenvector.

We use numerical integration techniques to construct the
overlap-integral matrix. It will have dim$(\pmb{I})$ eigenvectors.
Those corresponding to zero modes will not only be small, but will
have values that decrease quadratically with the (average) width of
the two-pion-mass bins, owing to the construction of the Gram matrix.

\begin{figure}[!t]
  \centering
  \includegraphics[width=0.9\columnwidth]{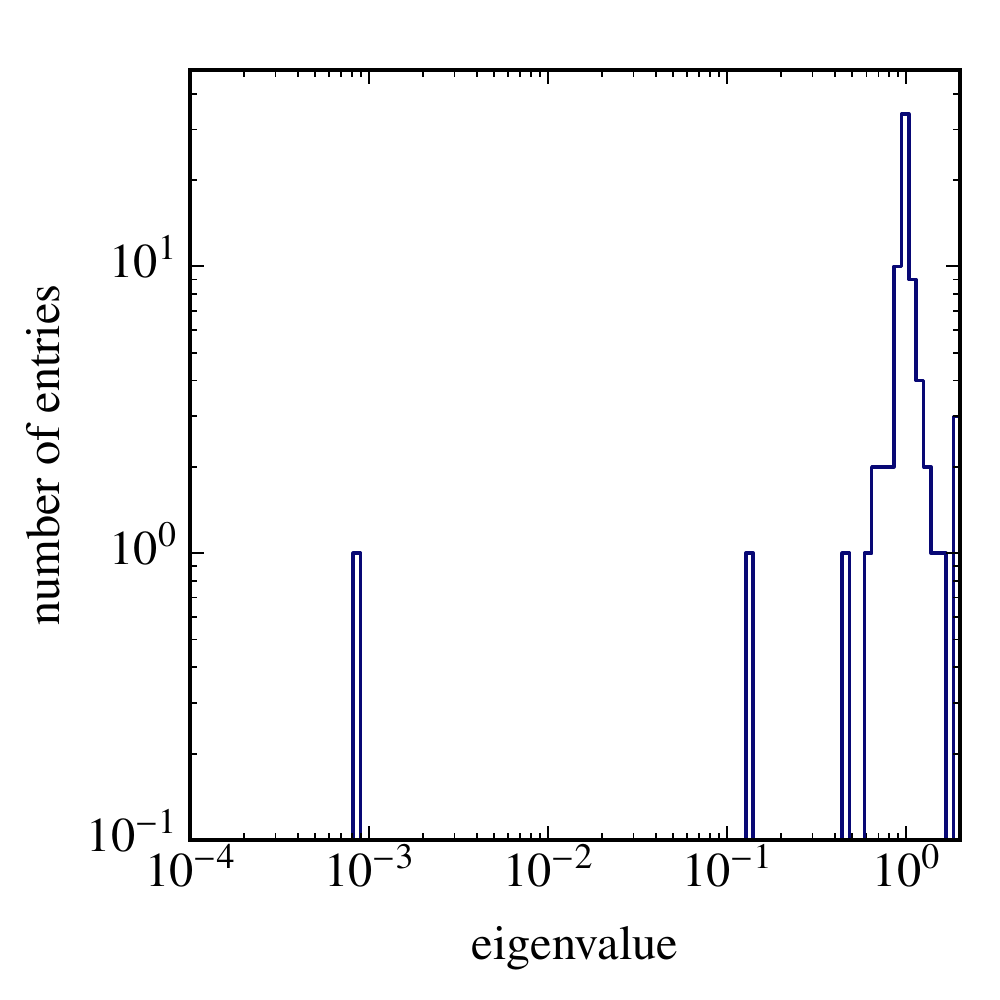}\\
  \includegraphics[width=0.9\columnwidth]{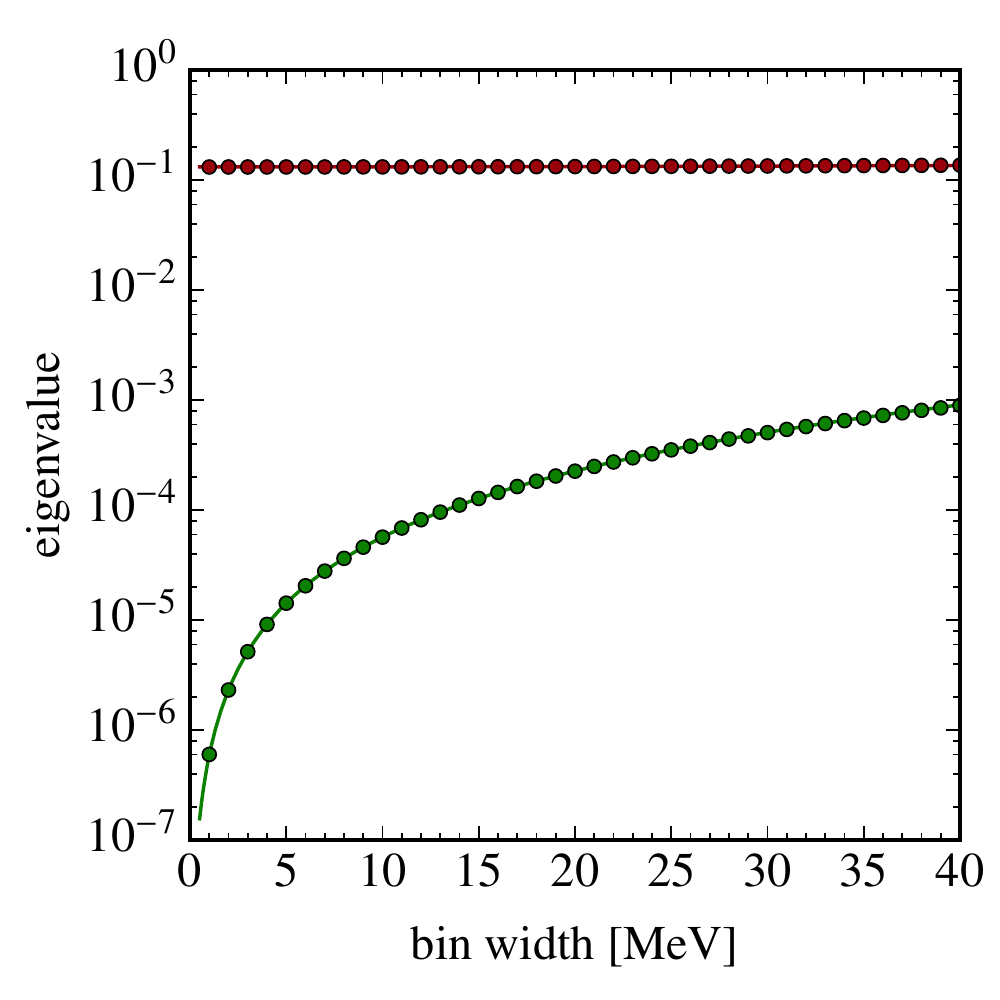}
  \caption{\label{fig:D3pi_num_zero_mode_eigen}
    Eigenvalue spectrum~(top) of the integral matrix for freed \Swave
    and \Pwave waves in \decay{\PDminus}{\Ppiminus\Ppiplus\Ppiminus};
    and the mass dependence of the two smallest eigenvalues~(bottom).}
  
\end{figure}

\begin{figure}[!t]
  \centering
  \includegraphics[width=0.9\columnwidth]{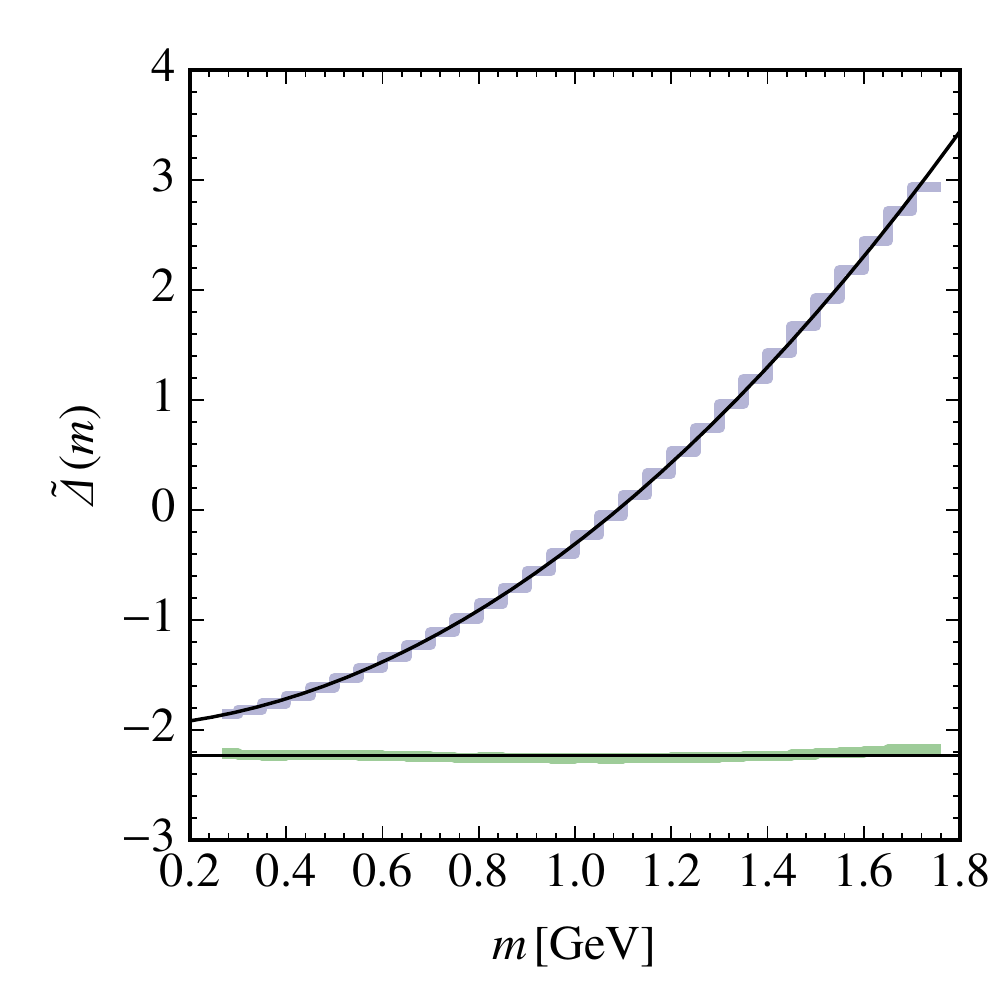}
  \caption{\label{fig:D3pi_num_zero_mode}
    Components of the numerically determined zero mode for freed
    \Swave~(blue) and \Pwave~(green) waves in
    \decay{\PDminus}{\Ppiminus\Ppiplus\Ppiminus} and the zero mode of
    equations~(\ref{eqn:sp_zero_mode_p})
    and~(\ref{eqn:sp_zero_mode_s})~(black).}
\end{figure}

We demonstrate this technique with the decay
\decay{\PDminus}{\Ppiminus\Ppiplus\Ppiminus} with the \Swave and
\Pwave waves freed. We expect to find the zero mode of
equations~(\ref{eqn:sp_zero_mode_p})
and~(\ref{eqn:sp_zero_mode_s}). The top plot in
figure~\ref{fig:D3pi_num_zero_mode_eigen} shows the eigenvalue spectrum of
the integral matrix. There is one significantly small eigenvalue. The
bottom plot in the figure shows the dependence of this eigenvalue and
the next-largest one on the average bin width. The smallest eigenvalue
quadratically depends on the bin width, but the next-largest one is
constant. Figure~\ref{fig:D3pi_num_zero_mode} shows the zero mode
formed from the eigenvector alongside that of
equations~(\ref{eqn:sp_zero_mode_p})
and~(\ref{eqn:sp_zero_mode_s})---the two are nearly
indistinguishable\footnote{The shapes of the zero mode also include
  barrier factors and form factors. If these are put into an analysis
  model explicitly, the shapes of the zero mode change
  accordingly.}. They deviate from each other only in the \Pwave-wave
high-mass region: The \Swave-wave analytical zero mode varies most
rapidly at high masses, so the step function is a worse approximation
there than it is at low masses. The \Pwave-wave step function at high
mass deviates from the expected form to compensate for the discrepency
in the \Swave wave. When we reduce the bin widths, this discrepency
disappears.

\section{Resolving zero-mode ambiguities}
\label{sec:resolving_ambiguities}

When we fit a freed-isobar PWA model to data, zero modes will
contribute to the dynamic amplitudes the fitter finds---the
$\miFitAmp^{a}_{\beta}$ of
equation~(\ref{eqn:model_independent_amp})---with their complex
$\zmAmp_\zmVec$ taking on arbitrary values as artifacts of the fitting
process. We can correct for their presence using knowledge of their
shapes and some assumption about the true underlying amplitudes and
recover the underlying physical values, the $\miTrueAmp^{a}_{\beta}$,
that are the goal of an analysis.  We refer to the fit that determines
the $\miFitAmp^{a}_{\beta}$ as the fit to data and further steps
(including additional fits) that determine the
$\miTrueAmp^{a}_{\beta}$ as the zero-mode correction.

There are many possible assumptions we can make about the true
underlying physical amplitudes; we present three examples below, which
each assumes a model for some part of the freed isobar
amplitudes. This gives us expectations for the underlying physical
values, which we label
$\miExpAmp^{a}_{\beta}$. Equation~(\ref{eqn:mi_amp_components}) tells
us that the difference between the underlying value and our
expectation is
\begin{equation}
  \miTrueAmp^{a}_{\beta} - \miExpAmp^{a}_{\beta} = \miFitAmp^{a}_{\beta} - \sum_{\zmVec} \zmAmp_{\zmVec} \zmVec^{a}_{\beta} - \miExpAmp^{a}_{\beta};
\end{equation}
since we can learn the $\zmVec^{a}_{\beta}$ using the method described
in section~\ref{sec:numerical_zero_modes}, the only unkowns are the
$\zmAmp_{\zmVec}$. We can fit for the $\zmAmp_{\zmVec}$ and correct
for them, yielding the true underlying values. We do this by
minimizing
\begin{widetext}
  \begin{equation}
    \label{eqn:chi2_def}
    \chi^2 \equiv \sum_{a\beta,b\delta}
    \qty(\miFitAmp^{a}_{\beta} - \sum_\zmVec \zmAmp_\zmVec \zmVec^{a}_{\beta} - \miExpAmp^{a}_{\beta})
    C^{-1}_{a\beta,b\delta}
    \qty(\miFitAmp^{b}_{\delta} - \sum_\zmVec \zmAmp_\zmVec \zmVec^{b}_{\delta} - \miExpAmp^{b}_{\delta}),
  \end{equation}
\end{widetext}
where $a$, $\beta$, $b$, and $\delta$ are as defined for
equation~(\ref{eqn:normalized_integral_matrix}) and
$C_{a\beta,b\delta}$ is the matrix of covariances of the
$\miFitAmp^{a}_{\beta}$ determined by the fit to data.

This step is not equivalent to having assumed a model from the very
start: The model-independent approach has many more degrees of freedom
than model-dependent approaches. The zero-mode correction step only
reduces the number of degrees of freedom in the analysis by the number
of free parameters in equation~(\ref{eqn:chi2_def}), leaving still
much more freedom than in a model-dependent analysis.

\subsection{Zero-mode correction examples}
\label{sec:zero_mode_correction_examplesx}

We demonstrate zero-mode correction using simulated data of three-pion
states produced by \PD-meson decay and pion--proton scattering. We
test three types of assumptions to correct for the zero mode: a model
that predicts a value in every two-pion-mass bin in every freed wave
with the $\zmAmp_\zmVec$ the only free parameters; a model for only a
subset of bins, with the $\zmAmp_\zmVec$ the only free parameters; and
a model with additional free parameters beyond those of the zero mode.

For \decay{\PD}{\Ppiminus\Ppiplus\Ppiminus}, we generated one million
events according to a model containing $\Pfzero(980)\Ppiminus$ in the
\Swave wave and $\Prho(770)\Ppiminus$ in the \Pwave wave, with both
resonances modeled by the relativistic Breit-Wigner
lineshape~\cite{PhysRev.49.519}. For the spin-dependent amplitudes, we
used the Zemach formalism of~\cite{PhysRev.140.B97}.

Our \PD-decay fit model frees both the \Swave and \Pwave waves. The
steps in both freed waves are contiguous and cover the full mass range
from $2m_{\Ppi}$ to $m_{\PD}-m_{\Ppi}$; they are 20~MeV wide. With
freed \Swave and \Pwave waves, as we demonstrated above, there will be
a zero mode with one complex degree of freedom.

For \decay{\Ppiminus\Pproton}{\Ppiminus\Ppiplus\Ppiminus\Pproton}, we
generated 260,000 events in the three-pion mass range from
\SI{1.50}{GeV} to \SI{1.54}{GeV} according to the model used by the
\compass collaboration in~\cite{PhysRevD.95.032004} with the
parameters they extracted from data. We used the helicity formalism
of~\cite{AnnalsPhys.7.404, PhysRevD.48.1225, Richman:1984gh}, to be
consistent with the analysis in~\cite{PhysRevD.95.032004}. The
\compass model contains 88 partial waves. Each is a unique combination
of quantum numbers for the three-pion state, a dynamic amplitude model
for the two-pion isobar (denoted by \Pxi), and an orbital angular
momentum between the isobar and the spectator pion. The three-pion
quantum numbers are formulated in the reflectivity
basis~\cite{PhysRevD.11.633}, with spin projection $M$ and
reflectivity $\epsilon$. This is all abbreviated as
$J^{PC}M^\epsilon\Pxi\Ppi L$.

\begin{table}[!b]
  \centering
  \begin{tabular}{l | *{4}{l@{\hspace{0.5em}}}@{\hspace{0.25em}} | l}
    \toprule

    $J^{PC}M^\epsilon$ & \multicolumn{4}{l}{$\freed{j}{L}$} \vrule & \\
    \hline

    $0^{-+}0^{+}$ &
    \freed{\Swave}{\Swave},&
    \freed{\Pwave}{\Pwave}&
    && $\checkmark$\\

    $1^{++}0^{+}$ &
    \freed{\Swave}{\Pwave},&
    \freed{\Pwave}{\Swave}&
    && $\checkmark$\\

    $1^{++}1^{+}$ &&
    \freed{\Pwave}{\Swave}&
    &&\\
    
    $2^{-+}0^{+}$ &
    \freed{\Swave}{\Dwave},&
    \freed{\Pwave}{\Pwave},&
    \freed{\Pwave}{\Fwave},&
    \freed{\Dwave}{\Swave} &
    $\checkmark$ \\

    $2^{-+}1^{+}$ &&
    \freed{\Pwave}{\Pwave},&
    &\\

    $2^{++}1^{+}$ &&
    \freed{\Pwave}{\Dwave},&
    &\\

    \bottomrule
  \end{tabular}
  \caption{ Freed waves in
    \decay{\Ppiminus\Pproton}{\Ppiminus\Ppiplus\Ppiminus\Pproton},
    grouped by three-pion quantum numbers, with presence of a zero
    mode indicated in the last column. \label{tab:freed_waves}}
\end{table}

Our $\Ppi\Pproton$-scattering fit model frees the dynamic amplitudes
in eleven different waves distinguished by the quantum numbers of the
three-pion system; the spin, $j$, of the two-pion isobar, which we
write $\free{j}$; and the angular momentum between the isobar and the
spectator pion. These eleven freed waves, listed in
table~\ref{tab:freed_waves}, replace fifteen waves of the \compass
model: waves with freed~\free{\Swave} replace those with the broad
$\Ppi\Ppi$ \Swave~wave and the $\Pfzero(980)$ (and $\Pfzero(1500)$ in
the $0^{-+}0^+$ wave); waves with freed~\free{\Pwave} replace those
with the $\Prho(770)$; and the wave with freed~\free{\Dwave} replaces
one with the $\Pftwo(1270)$. The remaining 73 waves are included in
their model-dependent formulation. Any combination of
model-independent and model-dependent waves is possible within the
freed-isobar approach; here we have freed the most prominent
waves. The two-pion mass bins in each freed wave are contiguous and
cover the full mass range from $2m_{\Ppi}$ to $m_{3\Ppi} - m_{\Ppi}$;
near the regions of the $\Prho(770)$ and $\Pftwo(1270)$, they are
\SI{20}{MeV} wide; near the $\Pfzero(980)$, they are \SI{10}{MeV}
wide; and elsewhere they are \SI{40}{MeV} wide.

\begin{figure*}[!t]
  \centering
  
  \includegraphics[width=.9\columnwidth]{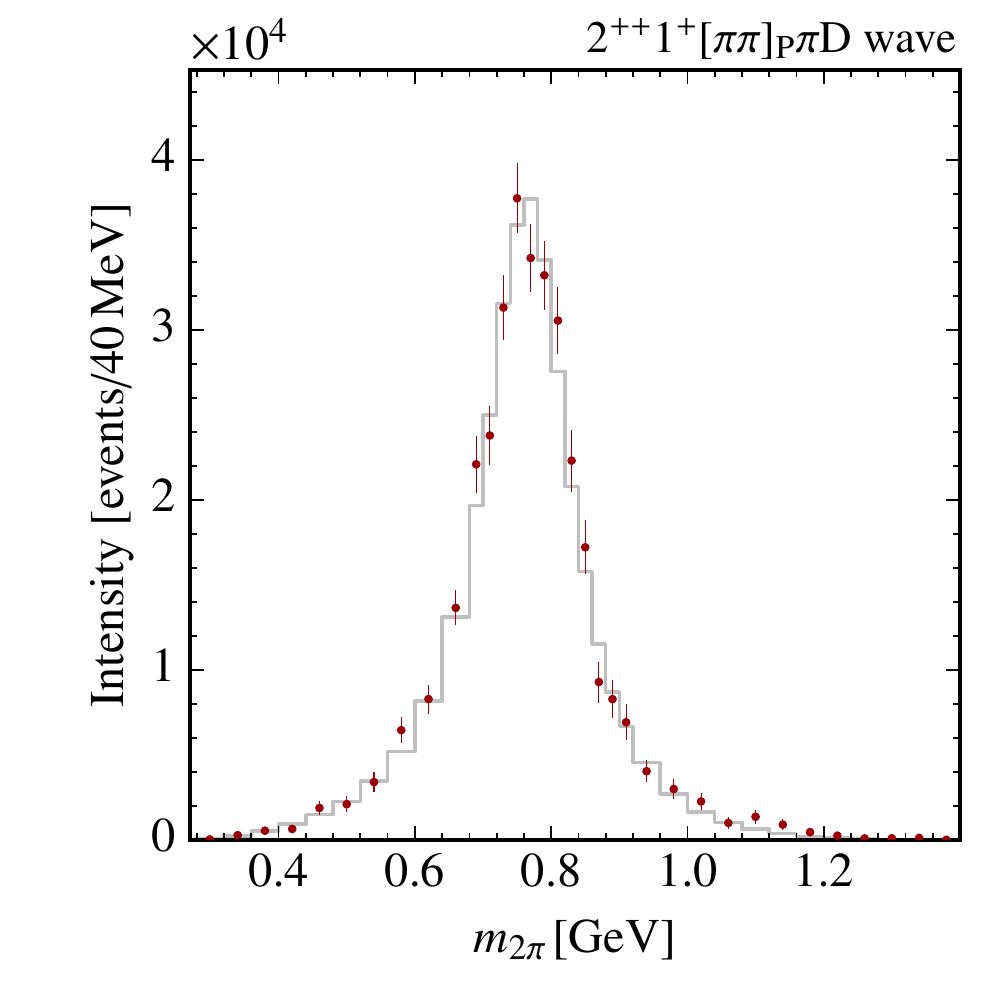}
  \hspace{\columnsep}
  \includegraphics[width=.9\columnwidth]{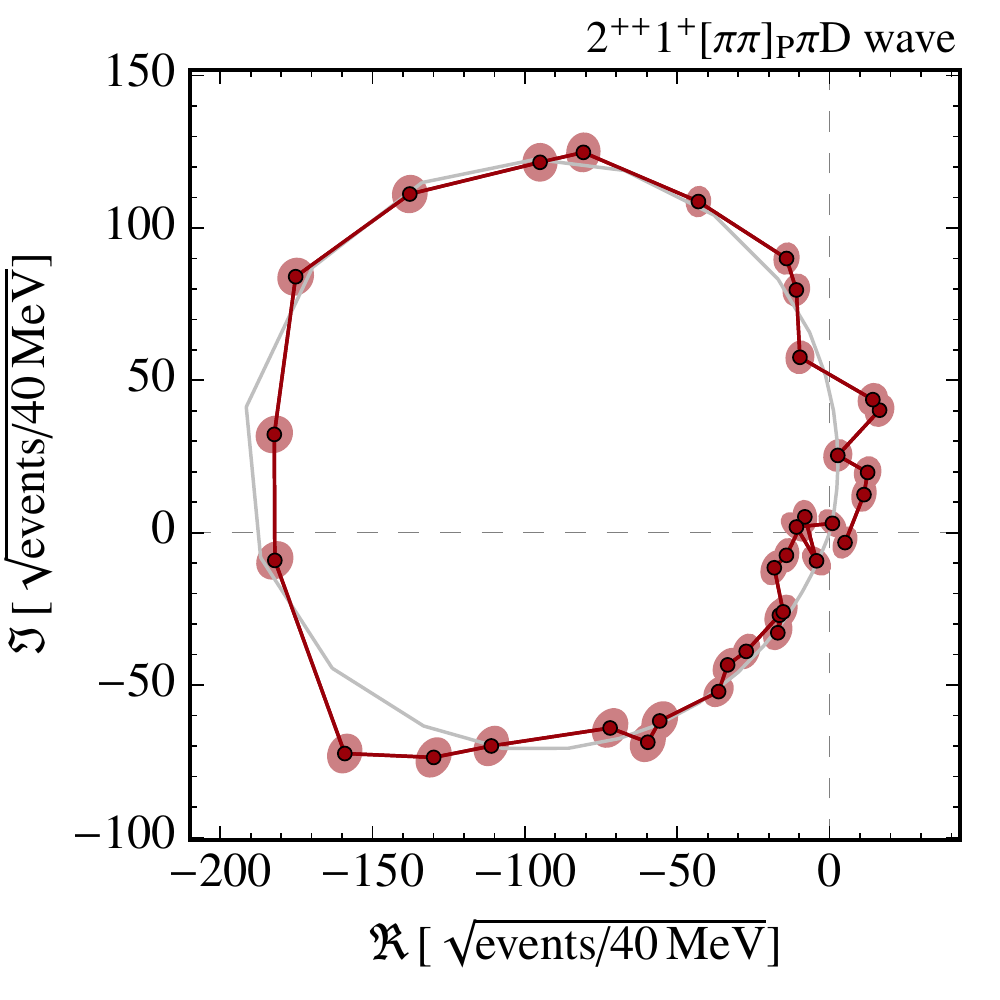}

  \caption{\label{fig:scattering_result_2pp} Intensity~(left) and
    dynamic amplitude in the complex plane~(right) of the simulated
    $\Ppi\Pproton$-scattering model~(grey) and the results of the fit
    to data~(red).}
\end{figure*}

Using the technique described in
section~\ref{sec:numerical_zero_modes}, we find the zero modes of this
scattering fit model. They do not connect waves with differing
initial-state quantum numbers. We find three zero modes in this
model---they are indicated in table~\ref{tab:freed_waves}; each has
one complex degree of freedom.

Figure~\ref{fig:scattering_result_2pp} shows an example result from
fitting this model to simulated data. The fit result is shown in red
and the generating model in gray. The left plot shows the intensity as
a function of two-pion mass; and the right plot shows the dynamic
amplitude in the complex plane. All other plots in this section are
identically structured. The freed-isobar shown is that of the
$2^{++}1^+\freed{\Pwave}{\Dwave}$ wave, which has no zero mode. The
fit result agrees well with the data-generation model.

\subsubsection{Complete-model constraint}

\begin{figure*}[!t]
  \centering
  
  \includegraphics[width=.9\columnwidth]{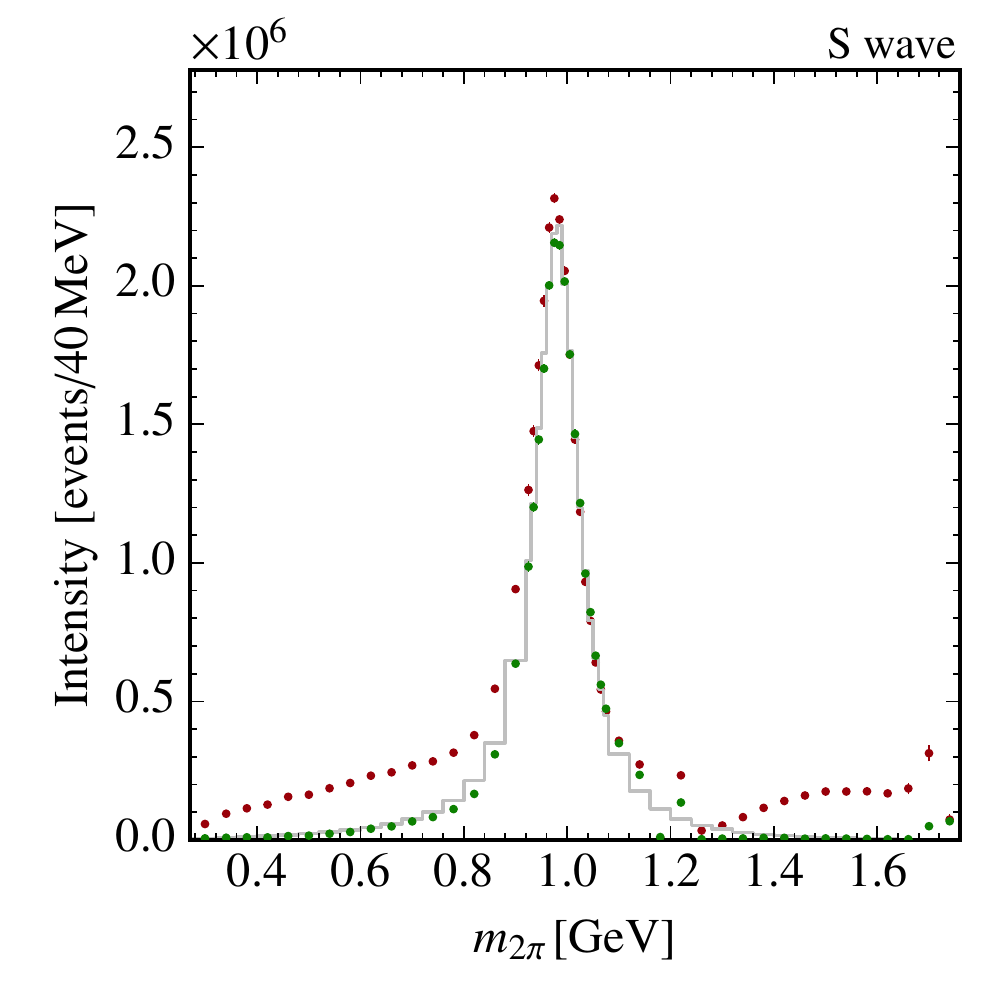}
  \hspace{\columnsep}
  \includegraphics[width=.9\columnwidth]{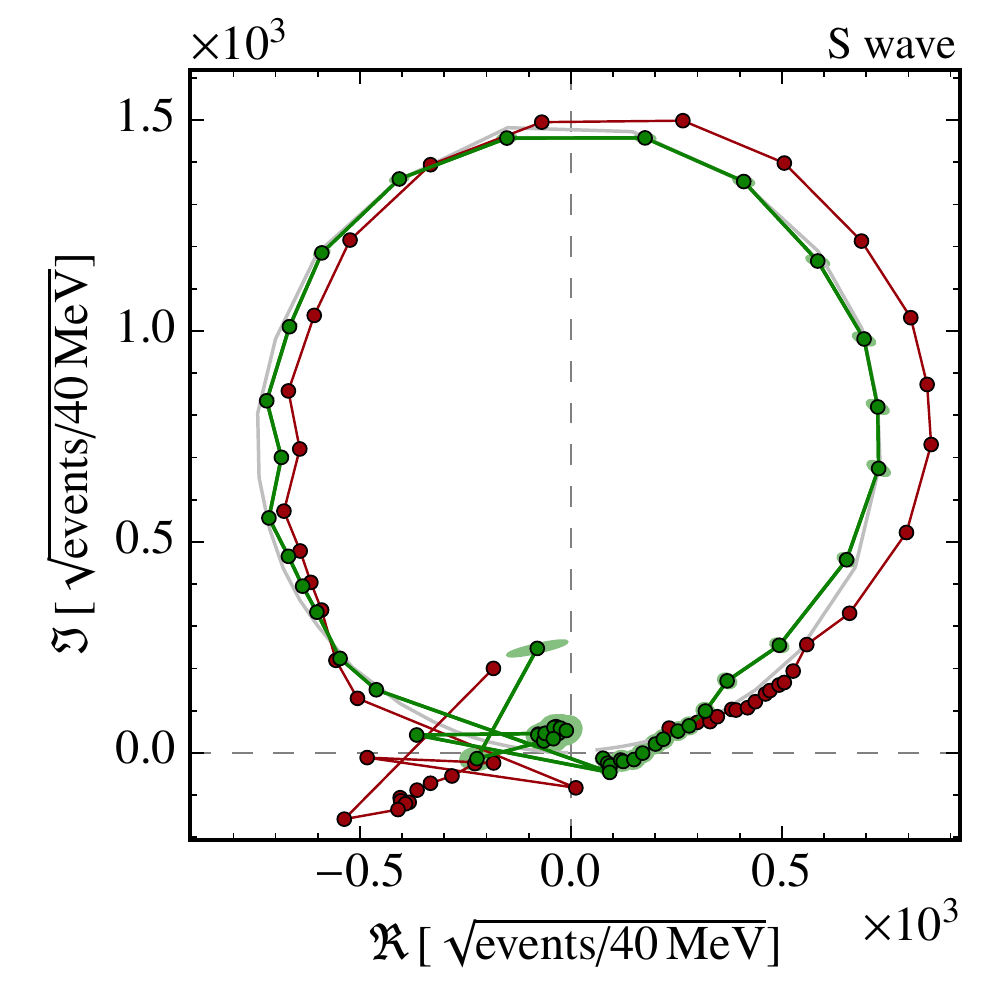}\\
  \includegraphics[width=.9\columnwidth]{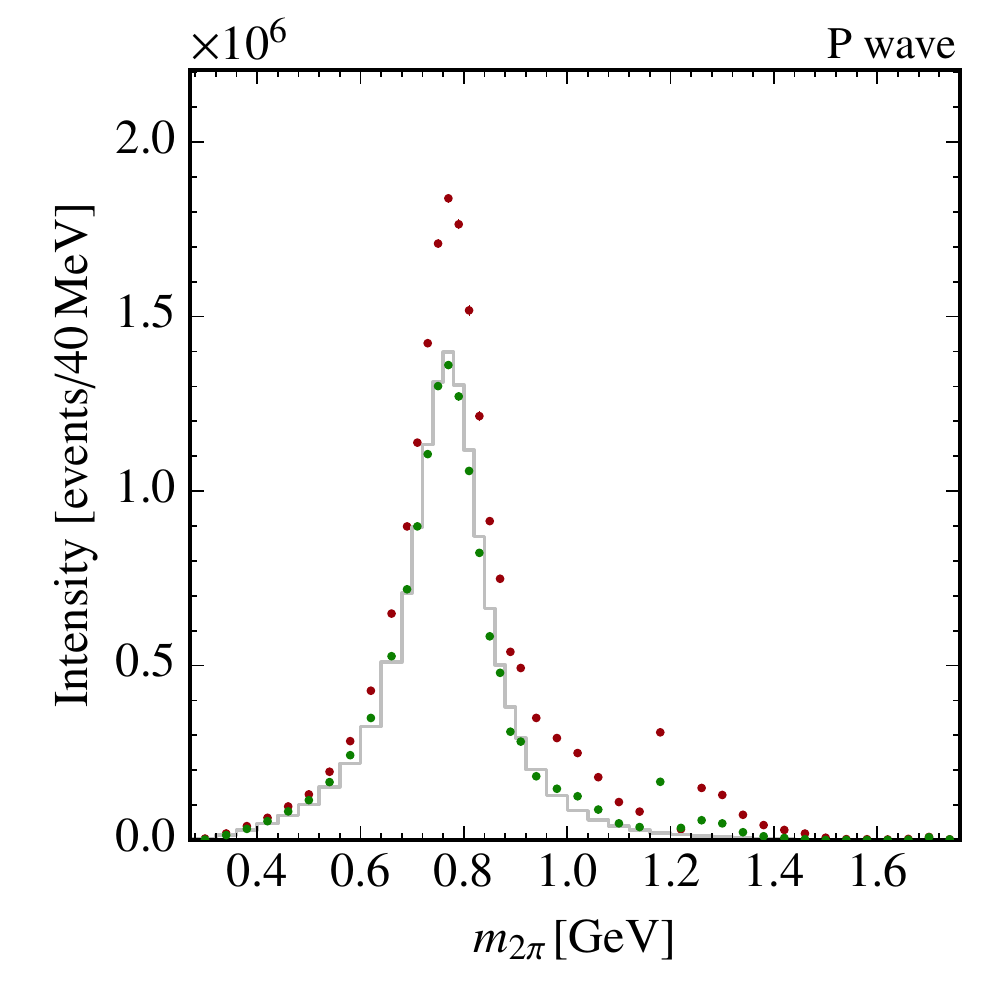}
  \hspace{\columnsep}
  \includegraphics[width=.9\columnwidth]{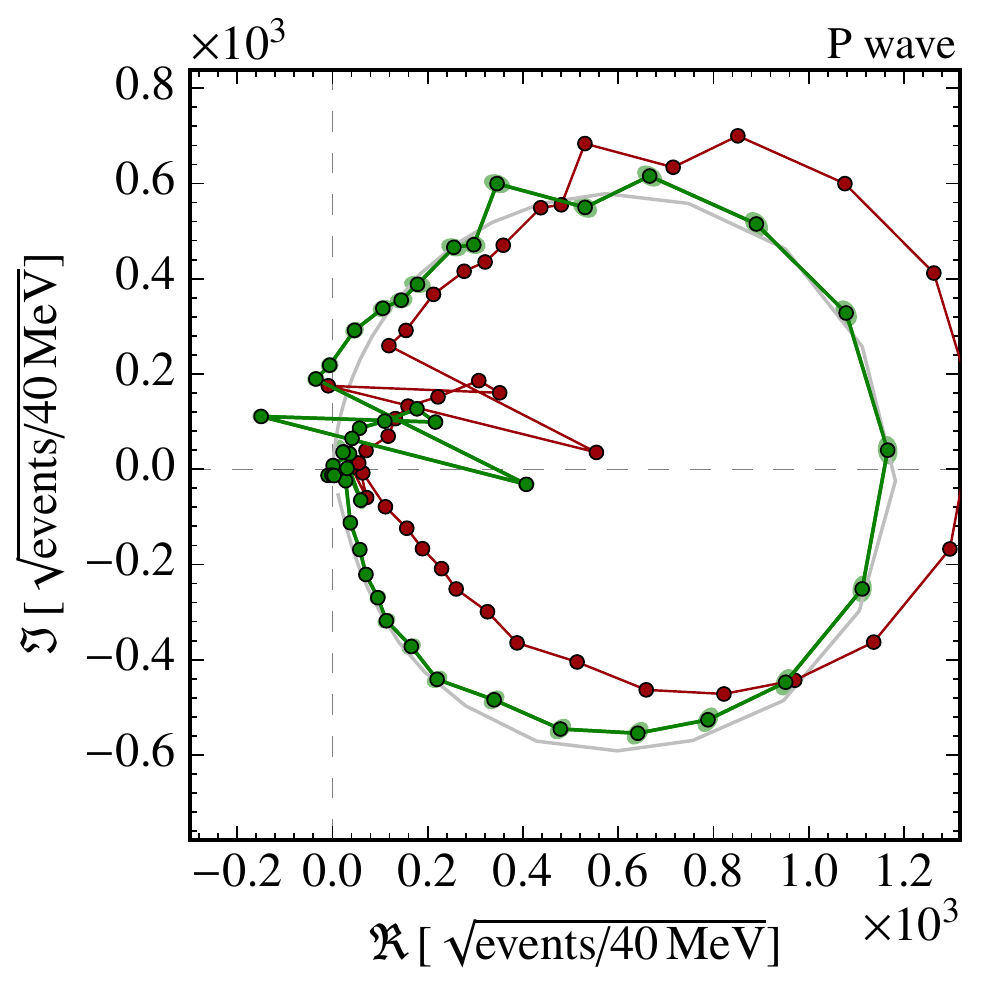}

  \caption{\label{fig:D_3pi_result} Intensities~(left) and dynamic
    amplitudes in the complex plane~(right) of the simulated
    \decay{\PDminus}{\Ppiminus\Ppiplus\Ppiminus} model~(grey), the
    results of the fit to data~($\miFitAmp^{a}_{\beta}$,~red), and the
    results after zero-mode
    correction~($\miTrueAmp^{a}_{\beta}$,~green).}
\end{figure*}

With our freed \PD-decay model, we determine the complex dynamic
amplitude value in each two-pion mass bin via a fit to simulated
data. In figure~\ref{fig:D_3pi_result}, the red points show the fit
results of the freed \Swave and \Pwave~waves. The grey lines show the
data-generation model. The fit result has peaks corresponding to those
of the generating model, but also considerable intensity elsewhere
that does not match the generating model.

To correct for the zero mode, we fit for its complex parameter,
\zmAmp, by minimizing the $\chi^2$ of
equation~(\ref{eqn:chi2_def}). We predict $\miExpAmp^{a}_{\beta}$ in
all bins of both the \Swave and \Pwave waves using a model that
contains both the $\Pfzero(980)$ and $\Prho(770)$---that is, our
original data-generation model. However, for our prediction we change
the masses and widths of the resonances: for the $\Pfzero(980)$, we
shift the mass from \SI{980}{MeV} to \SI{1}{GeV} and the width from
\SI{100}{MeV} to \SI{110}{MeV}; for the $\Prho(770)$, we shift the
mass from \SI{770}{MeV} to \SI{750}{MeV} and the width from
\SI{160}{MeV} to \SI{180}{MeV}.

In figure~\ref{fig:D_3pi_result}, the green points show the freed
waves with the zero mode subtracted given the value of \zmAmp found in
the second fit:
\begin{equation}
  \miTrueAmp^a_\beta = \miFitAmp^a_\beta - \zmAmp \zmVec^a_\beta.
\end{equation}
This result very closely resembles the generating model. Though we
predicted the $\miExpAmp^{a}_{\beta}$ with shifted values for the
masses and widths, our final result recovers the correct values. This
demonstrates that we do not need detailed and accurate expectations
for the zero-mode correction; nor can we coax a result out of the fit
via our expectation. This is in contrast to model-dependent PWA, which
is very sensitive to the fit model. However, though our expectations
for the zero-mode correction need not be detailed or accurate, they
must still be reasonable: We must predict a feature that the data in
some rough way contains. For example, we cannot predict
$\miExpAmp^{a}_{\beta}$ from a model of a resonance for which our data
is far below the threshold to produce since its features will be very
weak in the data.

\subsubsection{Partial-model constraint}

\begin{figure*}[!t]
  \centering
  
  \includegraphics[width=.9\columnwidth]{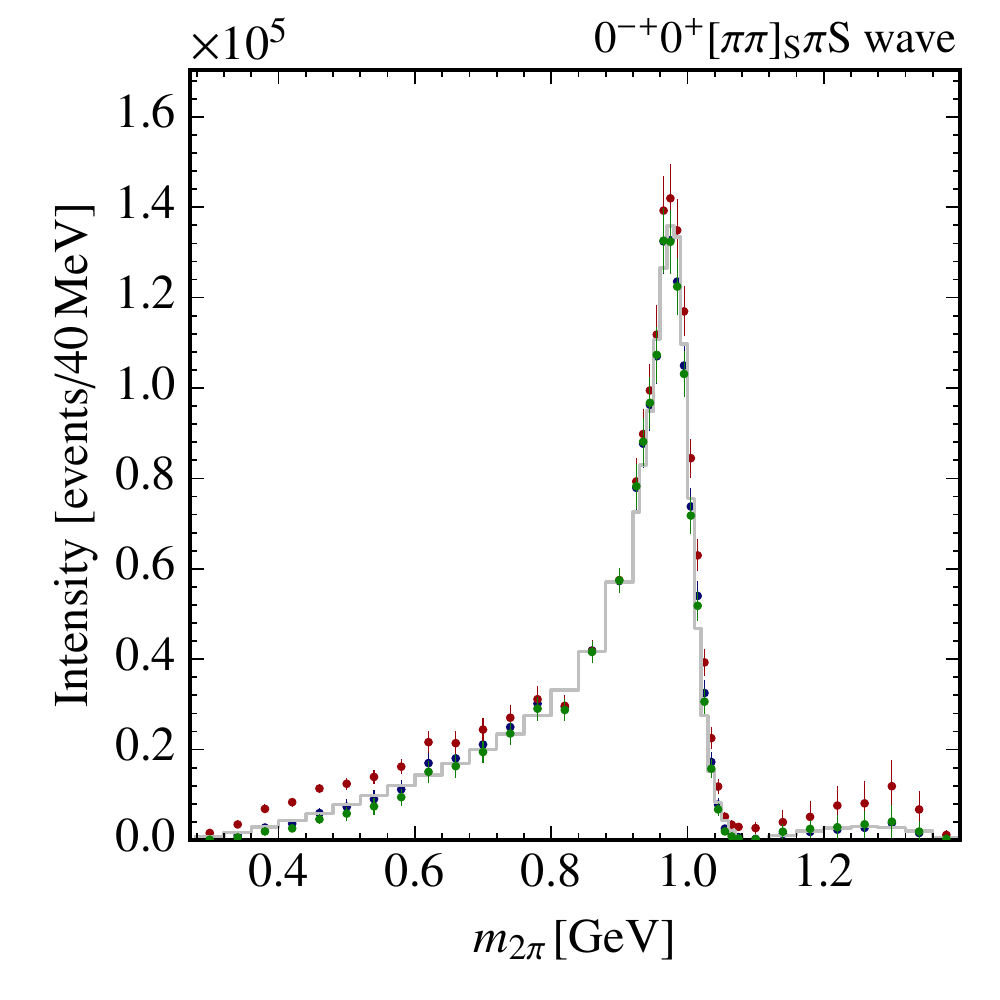}
  \hspace{\columnsep}
  \includegraphics[width=.9\columnwidth]{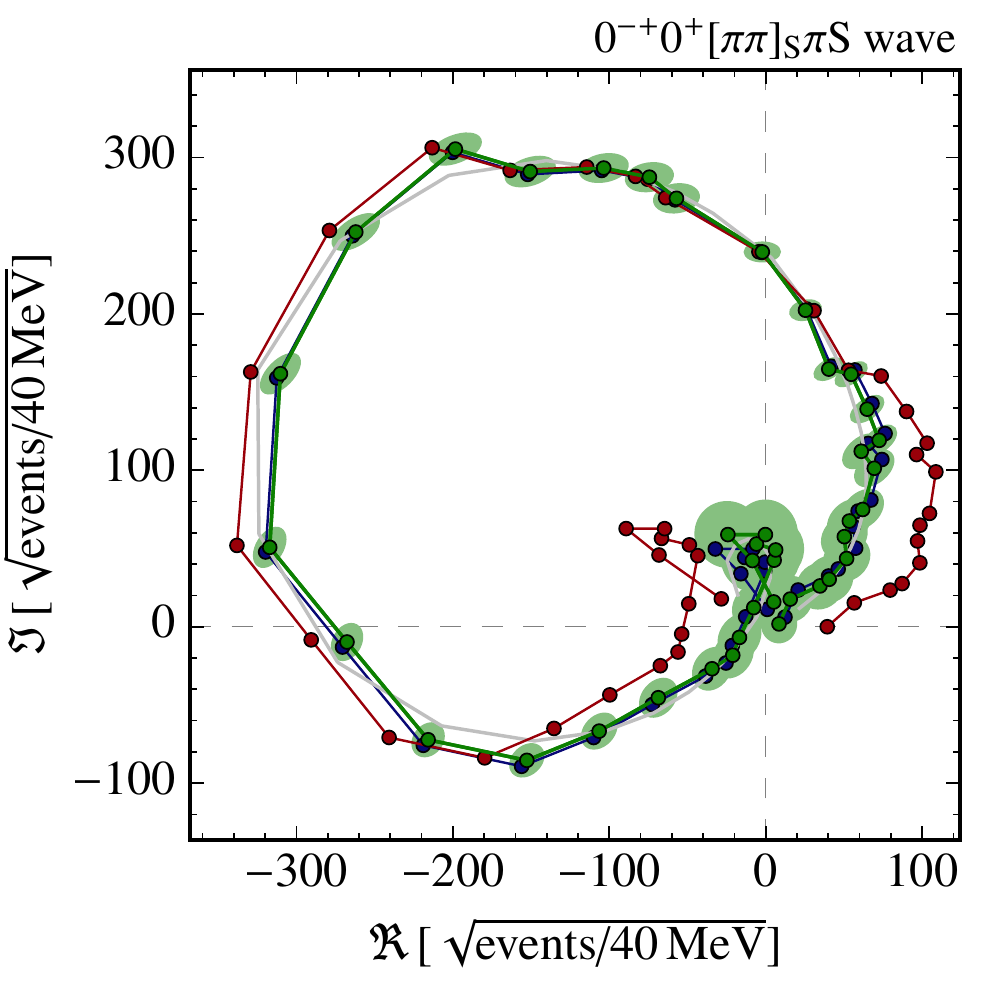}\\
  \includegraphics[width=.9\columnwidth]{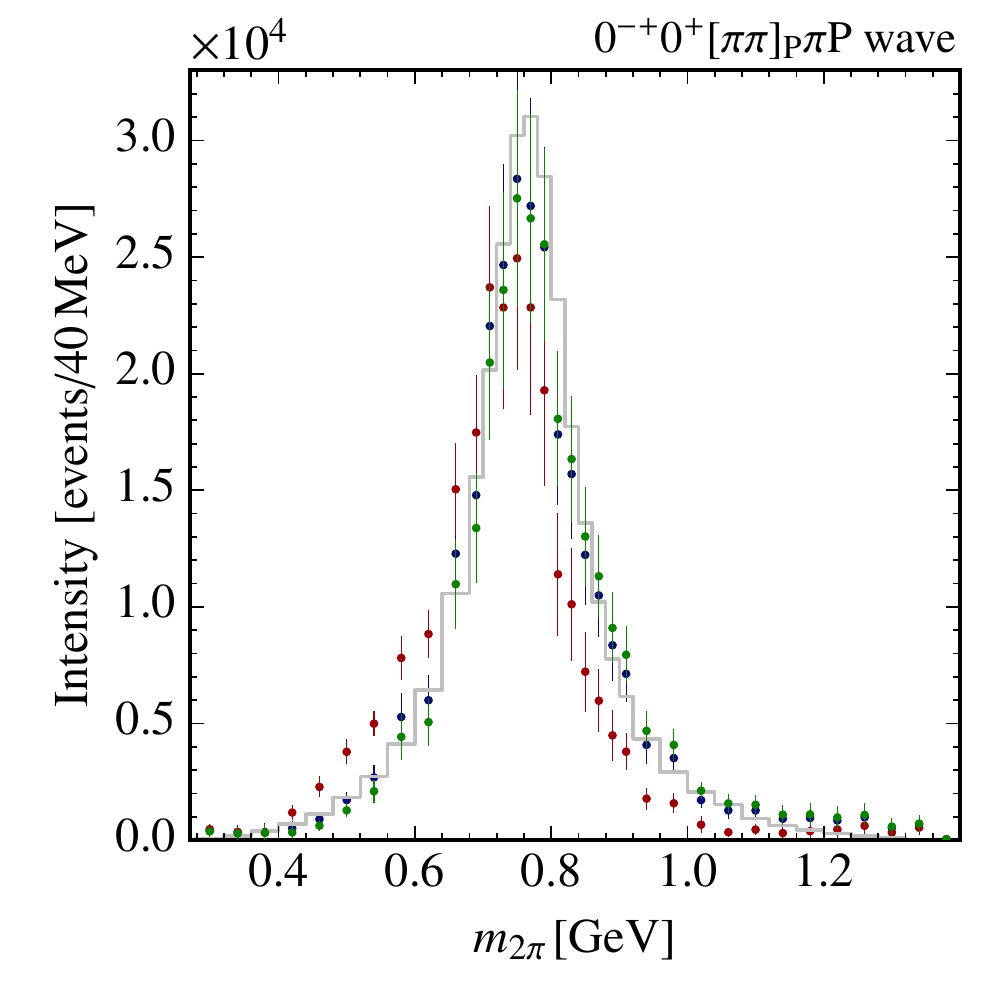}
  \hspace{\columnsep}
  \includegraphics[width=.9\columnwidth]{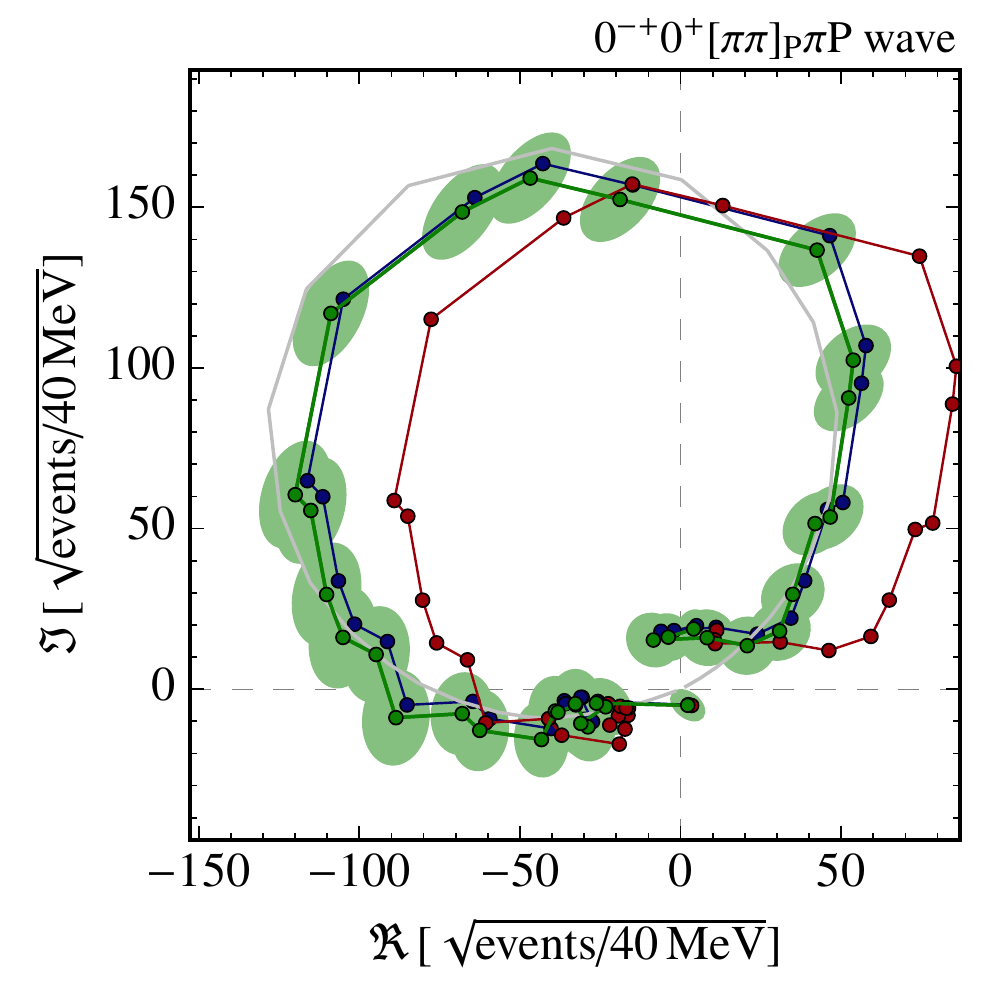}

  \caption{\label{fig:scattering_result_0mp} Intensities~(left) and
    dynamic amplitudes in the complex plane~(right) of the simulated
    $\Ppi\Pproton$-scattering model~(grey), the results of the fit to
    data~($\miFitAmp^{a}_{\beta}$,~red), and the results after
    zero-mode correction~($\miTrueAmp^{a}_{\beta}$) using both the
    \Swave and \Pwave waves~(blue, without uncertainties) and using
    only the \Pwave wave~(green).}
\end{figure*}

With our freed $\Ppi\Pproton$-scattering model, we determine the
complex dynamic amplitude value in each two-pion mass bin via a fit to
simulated data. In figure~\ref{fig:scattering_result_0mp}, the red
points show the fit results of the $0^{-+}0^+\freed{\Swave}{\Swave}$
and $0^{-+}0^+\freed{\Pwave}{\Pwave}$ freed waves. The grey lines show
the data-generation model. The disagreement between the fit result and
the generating model is due to the zero mode.

Again, to correct for the zero mode, we fit for its complex parameter,
\zmAmp, by minimizing $\chi^2$. Since the zero modes in our freed
$\Ppi\Pproton$-scattering model are contained entirely in waves with
the same initial-state quantum numbers, we need not assume a model for
the entire process---which, with 84 partial waves, is very
complicated---but only for the waves in which the zero mode arises. We
need only predict $\miExpAmp^{a}_{\beta}$ for the relevant waves, and
accordingly only sum over those waves in
equation~(\ref{eqn:chi2_def}). In
figure~\ref{fig:scattering_result_0mp}, the blue points show the freed
waves with the zero mode subtracted given the value of \zmAmp found in
the second fit. They agree well with the generating model.

Since the zero mode links the \freed{\Swave}{\Swave} wave and the
\freed{\Pwave}{\Pwave} wave, it is even enough to fit for \zmAmp in
only one of them. We restrict our prediction of
$\miExpAmp^{a}_{\beta}$ further to only the mass bins of the
\freed{\Pwave}{\Pwave} wave, and accordingly sum only over those bins
in equation~(\ref{eqn:chi2_def}). In
figure~\ref{fig:scattering_result_0mp}, the green points show
\emph{both} freed waves with the zero mode subtracted given the value
of \zmAmp found in a fit to \emph{only} the \Pwave~wave. The result very
closely resembles the generating model and agrees well with the
zero-mode correction that used both waves. This demonstrates that one
can correct for a zero mode with only a minimal assumption about a
model.

\subsubsection{Constraints with additional parameters}

In both zero-mode corrections above, the only free parameters were the
\zmAmp. Additional parameters used to predict the
$\miExpAmp^{a}_{\beta}$, such as masses and widths of resonances in
our assumed model, were fixed. But a common goal of PWA is to measure
such parameters. It would be superfluous and error prone to first
correct for the zero mode and then fit the zero-mode-corrected results
for such parameters. Instead, we should fit for them and the
contribution of the zero mode simultaneously.

We demonstrate this with a zero-mode-correction fit to the results in
the $0^{-+}0^{+}\freed{\Pwave}{\Pwave}$ wave of the
$\Ppi\Pproton$-scattering model (as determined, again, from a fit to
simulated data). We fit for both \zmAmp and the mass and width of the
$\Prho(770)$ (in a Breit-Wigner lineshape) that is contained in this
wave.  We recover a zero-mode corrected result identical to that shown
already in figure~\ref{fig:scattering_result_0mp}; and we find
\begin{alignat}{1}
  m_{\Prho} &= \SI{766.7 +- 1.8}{MeV}\\
  \Gamma_{\Prho} &= \SI{154.6 +- 4}{MeV},
\end{alignat}
which agree within their uncertainties with the simulated values
\SI{769.0}{MeV} and \SI{150.9}{MeV}.

\section{Conclusion}
\label{sec:conclusions}

Freed-isobar partial-wave analysis allows us to overcome the limits of
model-dependent analysis by using empirical step functions to describe
dynamic amplitudes. It can be useful for light-meson spectroscopy and
analysis of heavy-meson decays and hadronic \Ptau-lepton decays. In
particular, for CP-asymmetry measurements, freed-isobar PWA could be a
robust alternative to the common schemes of measuring asymmetries in
bins of phase space. Using the technique, we could instead determine
asymmetries in mass bins in distinct projections of the isobar quantum
numbers.

Several analyses have used freed-isobar PWA in limited ways, both as a
central analysis tool and as a cross check of model-dependent
analyses. But more expanded use of the technique---to fit with many
freed waves, whether initially or through a bootstrapping
procedure---has failed to produce meaningful results. This is due to
the presence of zero modes and their arbitrary degrees of
freedom.

We have demonstrated how to correct for these zero modes and remove
abritrary degrees of freedom using minimal assumptions. And we have
provided examples using simulated data of three-pion production via
both pion--proton scattering and \PD-meson decay. Our enhanced
freed-isobar PWA techniques may be useful for determining two-body
dynamic amplitudes that are consistent across a large variety of
strong-interaction problems; for example, one could test models of
final-state interaction.

\section*{Acknowledgements}

We acknowledge the support of the Cluster of Excellence Universe,
Exc153, Computational Centre for Astro- and Particle
Physics~(C$^2$PAP), Transregional Research Centre TR110, all funded by
the Deutsche Forschungsgemeinschaft. We would like to thank
M.~Pennington~(Sussex), A.~Szepaniak~(U.~Indiana), J.~Pelaez~(Madrid),
and W.~Ochs~(MPP) for many discussions in the course of this project.

\appendix
\section{Example Zero Modes}

The solutions to equation~\ref{eqn:zero_mode_functions}---the zero
modes---are dependent on the formalism chosen for $\psi_a$ and are not
guaranteed to exist for all scattering or decay processes. The zero
modes are also dependent on the symmetrizations summed over---the
presence (or lack thereof) of identical particles in the final state.

To demonstrate conditions under which zero modes appear, we consider,
first, decay of a spinless meson to three spinless mesons, which is
important for heavy-meson decay; and then give an example in decay of
a spinful state.

\subsection{Zero modes in decays of spinless mesons}

Let us consider the decay
\begin{equation}
  \decay{\PX}{\Phs1\Phs2\Phs3},
\end{equation}
where all particles are spinless. For $\psi_a$, as in
section~\ref{sec:zero_mode_example_case}, we use the Zemach tensor
formalism. We limit our discussion to the \Swave and
\Pwave~waves. Since the initial state is spinless, the spin of the
resonance is always the same as the total orbital and spin angular
momenta of the wave. Therefore there is only one wave for each isobar
spin. For the decay to an isobar formed by $ij$ with spin $\ell$ and a
spectator final-state particle $k$ (with $i$, $j$, and $k$ standing
for a cyclic permutation of \Phs1, \Phs2, and \Phs3), the
spin-dependent amplitude is
\begin{equation}
  \label{eqn:zemach_general_ij}
  \psi^\ell_{ij}(m_X, \vec\tau) = |\vec{p}_i|^{\ell}\ |\vec{p}_k|^\ell\, P_\ell\!\left(\hat{p}_i\cdot\hat{p}_k\right),\\
\end{equation}
where $P_\ell$ is the $\ell$'th-order Legendre polynomial and the
momenta are in the $ij$ rest frame.

\subsubsection{Zero modes purely in \Swave waves}
\label{subsubsec:swave_zero_mode}

The \Swave-wave spin-dependent amplitude is unity. Therefore the
\Swave-wave component of equation~(\ref{eqn:zero_mode_functions}) is
\begin{equation}
  \hat{\sum_a}\; \tilde\Delta_a^{\Swave}(\hat{m}).
\end{equation}
We can compensate a constant complex pedestal in one \Swave wave by
subtracting that same constant complex pedestal from any other \Swave
wave. Therefore, if there is more than one \Swave wave in the model,
there are zero modes that link each pair of \Swave waves. We can most
simply represent this with a set of zero modes that are constant in
each \Swave~wave,
\begin{equation}
  \label{eqn:Swave_zero_mode}
  \tilde\Delta_a^{\Swave}(m) = \tilde\beta_a,
\end{equation}
such that
\begin{equation}
  \label{eqn:Swave_zero_mode_constraint}
  \hat{\sum_a}\; \tilde\beta_a = 0,
\end{equation}
where the $\tilde\beta_a$ are complex variables, one per wave, each
with two real degrees of freedom; for $N_{\Swave}$ independent \Swave
waves, there are $(N_{\Swave}-1)$ free arbitary complex variables.

Such a zero mode arises, for example, in the decay
\decay{\PDminus}{\PKplus\PKminus\Ppiminus}, in which there are two
\Swave waves: in $\PKplus\PKminus$ and $\PKplus\Ppiminus$. In
contrast, no such zero mode arises in
\decay{\PDminus}{\Ppiminus\Ppiplus\Ppiminus} since there is only one
\Swave wave---in $\Ppiplus\Ppiminus$---with two symmetrizations. This
illustrates the difference between symmetrizations---the swapping of
identical final-state particles in and out of the isobar---and
waves---different groupings of particle species into
isobars. Different waves have independent dynamic amplitude; but
different symmetrizations of a single wave share a single dynamic
amplitude form.

\subsubsection{Zero mode purely in \Pwave waves}
\label{subsubsec:pwave_zero_mode}

The \Pwave-wave spin-dependent amplitude is
\begin{equation}
  \label{eqn:Zemach_Pwave}
  \psi^{\Pwave}_{ij}(\vec\tau) = \frac14 \qty( m_{jk}^2 - m_{ik}^2 - \qty(m_{\PX}^2-m_k^2)\frac{m_i^2-m_j^2}{m_{ij}^2} ).
\end{equation}
If each final-state particles is unique, there is no symmetrization
necessary and the \Pwave-wave component of
equation~(\ref{eqn:zero_mode_functions}) is
\begin{widetext}
  \begin{equation}
    \label{eqn:pwave_zero_mode}
    \frac{1}{4} \sum_{ijk}
    \qty(\tilde\Delta_{ik}^{\Pwave}(m_{ik}) - \tilde\Delta_{jk}^{\Pwave}(m_{jk})) m_{ij}^2
    - \qty(m_{\PX}^2-m_k^2)\frac{m_i^2-m_j^2}{m_{ij}^2} \,\tilde\Delta_{ij}^{\Pwave}(m_{ij}),
  \end{equation}
\end{widetext}
where the sum is over cyclic permutations of $\Phs1\Phs2\Phs3$ as
$ijk$, and we have labeled each wave by the two final state particles
forming its \Pwave-wave isobar. This amplitude is zero if all three
isobar configurations are allowed and all dynamic amplitudes are
\begin{equation}
  \Delta_a^{\Pwave}(m) = \tilde\gamma m^2,
\end{equation}
with one arbitrary complex variable, $\tilde\gamma$, for all
amplitudes.

Such a zero mode arises, for example, in the decay
\decay{\PDzero}{\Ppiplus\Ppiminus\Ppizero}, in which all \Pwave-wave
isobars are possible: $\Ppiplus\Ppiminus$, $\Ppiplus\Ppizero$, and
$\Ppiminus\Ppizero$. No such zero mode arises in, for example,
\decay{\PDminus}{\PKplus\PKminus\Ppiminus}, if we disallow an isobar
in $\PKminus\Ppiminus$ because there are no doubly-charged mesons.

If two of the final-state particles are the same species and charge,
as we have in the example decay
\decay{\PDminus}{\Ppiminus\Ppiplus\Ppiminus}, there is no purely
\Pwave-wave zero mode.

\subsubsection{Zero modes connecting \Swave and \Pwave waves}
\label{subsubsec:spwave_zero_mode}

Let us extend the zero mode of equations~(\ref{eqn:sp_zero_mode_p})
and~(\ref{eqn:sp_zero_mode_s}) for the case of \Phs1, \Phs2, and \Phs3
each a unique particle species: The zero mode is constant in the
\Pwave waves, but now each combination of final-state particles has an
independent dynamic amplitude:
\begin{equation}
  \tilde\Delta^{\Pwave}_{ij}(m) = 4\zmAmp_{ij}.
\end{equation}
Substituting this into equation~(\ref{eqn:pwave_zero_mode}) gives the
total \Pwave wave amplitude:
\begin{equation}
  \sum_{ijk}\, \qty(\zmAmp_{ik} - \zmAmp_{jk}) \;\! m^2_{ij} - \zmAmp_{ij} \qty(m^2_{\PX} - m^2_k) \frac{m^2_i - m^2_j}{m^2_{ij}},
\end{equation}
where the sum, as above, is over cyclic permutations of
$\Phs1\Phs2\Phs3$ as $ijk$. The summand is dependent on only one
mass---the isobar mass. We can perfectly balance each term in the sum
with \Swave-wave dynamic amplitudes
\begin{equation}
  \tilde\Delta^{\Swave}_{ij}(m) = -\qty(\zmAmp_{ik} - \zmAmp_{jk})\;\! m^2  + \zmAmp_{ij} \qty(m^2_{\PX} - m^2_k) \frac{m^2_i - m^2_j}{m^2},
\end{equation}
and get a total amplitude of zero. Such a zero mode arises if all
\Swave waves and any \Pwave wave are freed.

\subsubsection{All zero modes in the \Swave and \Pwave waves}

\begin{figure*}[t]
  \centering
  
  \includegraphics[width=0.9\columnwidth]{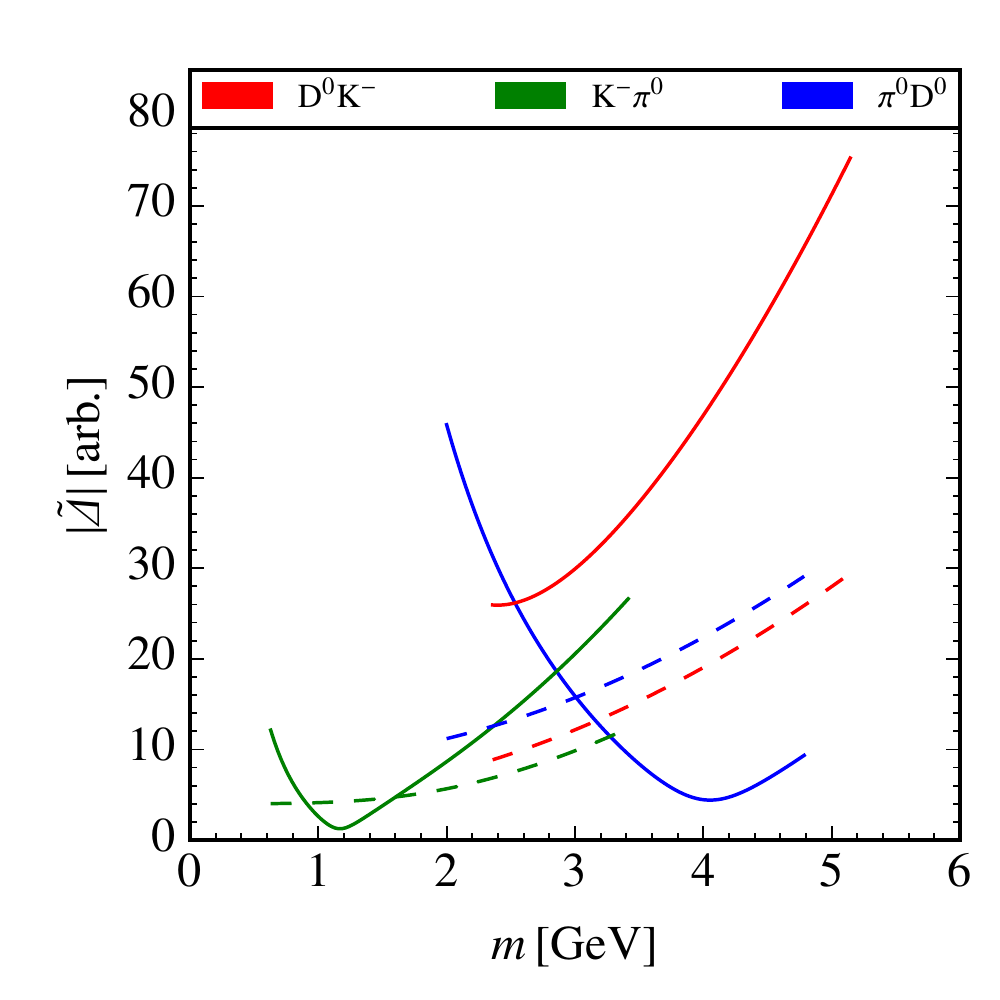}
  \hspace{\columnsep}
  \includegraphics[width=0.9\columnwidth]{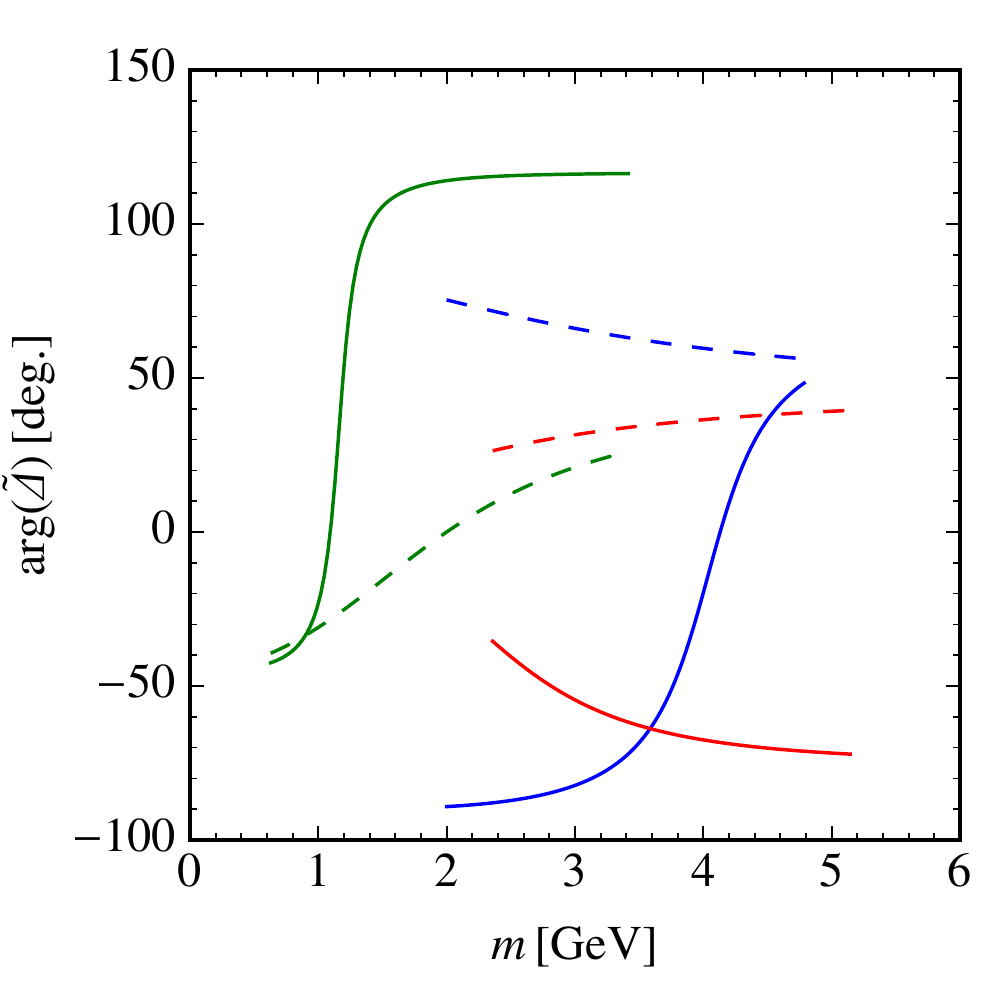}

  \caption{\label{fig:zero_mode_B_D0Kpi0}
    \Swave-wave (solid) and \Pwave-wave (dashed) dynamic components
    (magnitude, top; phase, bottom) of the generic zero mode in the
    decay \decay{\PBminus}{\PDzero\PKminus\Ppizero} with
    $\zmAmp_{\PD\PK}{\,=\,}1$,
    $\zmAmp_{\PK\Ppi}{\,=\,}\exp(i\ang{45})$, and
    $\zmAmp_{\Ppi\PD}{\,=\,}2\exp(i\ang{90})$;
    $\tilde\beta_{\PD\PK}{\,=\,}1$,
    $\tilde\beta_{\PK\Ppi}{\,=\,}\exp(i\ang{120})$, and
    $\tilde\beta_{\Ppi\PD}{\,=\,}\exp(i\ang{240})$; and
    $\tilde\gamma{\,=\,}\exp(i\ang{45})$.}
\end{figure*}

If all \Swave and \Pwave waves are freed, then the three different
forms of zero mode demonstrated above are present:
\begin{alignat}{1}
  \tilde\Delta^{\Swave}_{ij}(m) &= \tilde\beta_{ij} + (\zmAmp_{jk} - \zmAmp_{ik})\;\! m^2 + \zmAmp_{ij} (m^2_{\PX} - m^2_k) \frac{m^2_i - m^2_j}{m^2}\\
  \tilde\Delta^{\Pwave}_{ij}(m) &= 4\;\!\zmAmp_{ij} + \tilde\gamma m^2,
\end{alignat}
There are seven arbitrary complex constants with one constraint:
$\hat{\sum}_{ij}\tilde\beta_{ij} = 0$. The combined shapes of all zero
modes are complex functions that may contain phase motion that mimics
a resonance. Figure~\ref{fig:zero_mode_B_D0Kpi0} shows such a
situation for the example decay
\decay{\PBminus}{\PDzero\PKminus\Ppizero} with an example set of
complex parameters for the zero mode. Large phase motion manifests in
two of the waves---such a shape, if not corrected for, could lead to a
wrong interpretation of an analysis result.

\subsection{Further zero modes}

Zero modes are seen in many other combinations of isobars beyond the
above examples of spinless meson decays. One decay of particular
interest is that of a $1^{-+}$ state into a $1^{--}$ isobar and a
pseudoscalar meson in a relative \Pwave~wave with two symmetrizations
(as exists in our example final state,
$\Ppiminus\Ppiplus\Ppiminus$). We can write the amplitude using the
relativistic tensor formalism of references~\cite{JPhysG.28.15,
  PhysRevD.51.2247, PhysRevD.57.431} as
\begin{equation}
  \qty[\psi^{1^{-+}}_{(ij)k}(\vec\tau)]_\mu \propto \epsilon_{\mu\nu\rho\sigma} \, p_i^\nu \, p_j^\rho \, p_k^\sigma;
\end{equation}
note, that this amplitude is a vector, since it describes a spin-one
quantity. Because of the Levi-Civita tensor,~$\epsilon$, this changes
sign under exchange of two indices. The symmetrized amplitude is
therefore proportional to
\begin{equation}
  \epsilon_{\mu\nu\rho\sigma} \, p_1^\nu \, p_2^\rho \, p_3^\sigma \, \qty( \tilde\Delta(m_{12}) - \tilde\Delta(m_{23}) ).
\end{equation}
This is equal to zero everywhere if $\tilde\Delta(m) =
\tilde\Delta(m')$ for all $m$ and $m'$---that is, if $\tilde\Delta(m)$
is constant. Since $\tilde\Delta(m)$ may be complex, the zero mode has
two degrees of freedom entirely contained in one isobar.


\end{document}